# Exploring conformational energy landscape of glassy disaccharides by CPMAS $^{13}$C NMR and DFT/GIAO simulations. I. Methodological aspects.

Ronan LEFORT[1], Patrice BORDAT[2], Attilio CESARO[3], and Marc DESCAMPS[1]

[1]Laboratoire de Dynamique et Structure des Matériaux Moléculaires, P5, Université de Lille 1, Cité Scientifique, F-59655 Villeneuve d'Ascq Cedex, France

[2]Laboratoire de Chimie Théorique et de Physico-Chimie Moléculaire, UMR 5624 - IFR, LCS, 2, rue Jules Ferry, F-64000 Pau, France

[3]Laboratory of Physical and Macromolecular Chemistry, Department BBCM and UdR INSTM, University of Trieste, Via Giorgieri 1, I-34127 Trieste, Italy

Abstract :

The aim of this article is to assess the ability of chemical shift surfaces to provide structural information on conformational distributions of disaccharides in glassy solid state. The validity of the general method leading to a simulation of inhomogeneous $^{13}$C chemical shift distributions is discussed in detail. In particular, a proper consideration of extrema and saddle points of the chemical shift map correctly accounts for the observed discontinuities in the experimental CPMAS spectra. Provided that these basic requirements are met, DFT/GIAO chemical shift maps calculated on relaxed conformations lead to a very satisfactory description of the experimental lineshapes. On solid-state trehalose as a model of amorphous disaccharide, this simulation approach defines unambiguously the most populated sugar conformation in the glass, and can help in discriminating the validity of different models of intramolecular energy landscape. Application to other molecular systems with broad conformational populations is foreseen to produce a larger dependence of the calculated chemical shift distribution on the conformational map.

## Introduction

The ability of describing quantitatively oligosaccharide conformational distribution and flexibility in solution or in amorphous states has been for long time a challenge for computational methods (see for example ref. [1] and references therein). Refined force-fields and sophisticated computational approaches have been used time by time to evaluate the conformational energy landscape of flexible molecules. However, no clear cut could have been recognized about the general validity of the several force-fields available, and still very recently the question about whether conformation of simple sugars can be properly described is matter of debate [2, 3].

Going to larger systems, such as oligo- and poly-saccharides, the most relevant variables are undoubtedly the rotational angles about the glycosidic bonds, as long as dimensional and topological properties of these molecules mainly depend on these variables. A scrutiny of the results on different force-fields indicates an overall consensus about gross features of the contour energy landscape of disaccharide units, but the difference in the detail of the relative minima seems still unavoidably related to the specific model used. For long polymers, even more relevant is the fact that the use of force-fields based on small molecules has to be often stretched to reach the description of large molecules. More recently, for a hyaluronan chain the coherence with experimental data was reached by using the force field tuned with NMR relaxation data [4].

Indeed, some assessed prerogatives of computational methods have often been hampered by the difficulty of producing suitable experimental results, which however provided only thermodynamic averages, leaving unsolved the actual probability distribution. Thus, the most practical use of conformational energy surfaces resides either in a statistical (thermodynamic average) evaluation of experimentally measurable properties, or in the identification of the most probable conformational state associated to the molecular fragment. Whereas the statistical average and the most probable conformation are clearly associated to the solution and the crystalline state, respectively, the actual distribution of the conformational states and its relevance in the system properties has been investigated to a much less extent. The solution of this problem is directly related to the understanding of the conformational disorder in the amorphous (glassy) state.

With these problems in mind, exploitation is made in the present work toward the possibility of comparing simulated and experimental data on statistical conformational distributions by using $^{13}$C chemical shifts. It is now clearly established that $\sigma(\Omega)$ chemical shift and $E(\Omega)$ energy surfaces calculated in the phase space spanned by conformational degrees of freedom $\Omega$ are very efficient tools for understanding the observed $^{13}$C isotropic chemical shifts for carbohydrates [5-7]. In solution NMR studies, it is often assumed that molecular conformations are explored over time with a probability governed by $E(\Omega)$, and that the measured isotropic chemical shift is equal to the mean value obtained by averaging the $\sigma(\Omega)$ surface. This average can be either evaluated by a statistical model [6], or by exploring the $E(\Omega)$ map by molecular dynamics [7]. The accuracy of the results is therefore highly dependent on the validity and precision of three parameters: the Ramachandran map $E(\Omega)$, the chemical shift map $\sigma(\Omega)$, and the dynamical average.

In amorphous (glassy) carbohydrates, molecular conformations are statically distributed, and the CPMAS spectra often display heterogeneous line broadening, the chemical shift distributions reflecting the underlying conformational distributions [8-12]. Those broad lines often exhibit discontinuities and asymmetric features, and therefore provide much more information to be confronted to structural models, than a single averaged chemical shift value as measured in solution. Moreover, since dynamical average is not required in solid state, in principle the accuracy of calculations can be improved, this source of error vanishing. Hence, one can expect that the application of simulations based on the $\sigma(\Omega)/E(\Omega)$ maps could provide considerable insight on the actual local structure of glassy carbohydrates. Despite the large interest on the structural and dynamical features of sugar glasses and their practical implication in many fields, including the potential involvement in the bioprotection, to the best of our knowledge this kind of investigation is almost absent in literature. Indeed, the scarceness of such conformational solid state studies [5] is rather surprising, and might perhaps be ascribed only in part to the prohibitive CPU time required for all *ab initio* studies. Therefore, the main aim of this paper is to investigate the potentiality of chemical shift surfaces coupled to molecular mechanics conformational optimizations, applied to solid state NMR studies of glassy sugars.

## $^{13}$C Chemical shift calculations on disaccharides

**Experiments and methods**

*Trehalose samples.*

Amorphous (glassy) trehalose samples were prepared by ball-milling 1g of anhydrous trehalose ($T_\beta$) during 30 hours in a planetary grinder PULVERISETTE 7 (Fritsch, inc.). DSC thermograms were recorded after grinding in order to verify the total character of the amorphization via the glass transition and the subsequent amorphous state [8].

*NMR Experiments.*

The $^{13}$C CPMAS experiments were carried out at 100.6 MHz on a Bruker AV400 solid-state NMR spectrometer. Linear amplitude modulation of the rf field during the contact pulse (typ. 2 ms), and tppm heteronuclear decoupling during acquisition were employed. Recycle delays ranging between 200 and 450 s were used. Rotation speed was set to 5 kHz. A standard digital filter was used for acquisition, and the spectra were obtained by simple Fourier transform of the induction decay, without data apodization.

*Molecular modelling.*

Molecular modelling was achieved using the Hyperchem Pro 7 (Hypercube, Inc.) software package. All conformations were generated by constraining two dihedral angles $\phi$ and $\psi$. For trehalose, $\phi=(O_5-C_1-O_1-C'_1)$ and $\psi=(C_1-O_1-C'_1-O'_5)$, with the convention of $\phi=180°$, $\psi=180°$ for trans conformation. For fixed $\phi$ and $\psi$, all other conformational degrees of freedom were optimized using the molecular mechanics package provided in Hyperchem, including equivalent implementations of the AMBER and CHARMM empirical force fields. The parameters versions used will be denoted in the text by BIO85 (Reiher, [13]), CHARMM27 (Foloppe and MacKerell, [14]) and AMBER99 (Cornell, [15]). In order to take into account a mean-field contribution of the surrounding molecules in a condensed matter sample, a standard screening procedure of coulomb interaction was used, through an effective scaled dielectric constant. The 1-4 Scale Factors were equilibrated between Coulomb and Van der Waals interactions.

*Conformational energy maps.*

All calculations were started from the molecular coordinates obtained by x-ray structure determination on single crystal [16]. The starting crystal structure was relaxed by molecular mechanics in order to identify the lowest energy molecular conformation (LMC) in a given force field. For each different force field here used, an adiabatic Ramachandran map $E(\phi,\psi)$ was calculated by classically mapping the whole $(\phi,\psi)$ space by 324 (=18 x 18) points separated by 20° steps [5-7]. For each $(\phi,\psi)$ point in the map, minimization started from the LMC coordinates, then the $\phi$ and $\psi$ values were set and restrained by setting a high value of dihedral angle spring constants. All other degrees of freedom were then relaxed using the same force field parameters as for determining the LMC.

*Isotropic $^{13}C$ chemical shift maps.*

NMR chemical shift calculations were carried out for each studied disaccharide on the previous 324 relaxed conformations generated by molecular mechanics. For each conformation, the isotropic magnetic shielding of each carbon of the sugar was evaluated using the GIAO method on a density functional theory (DFT) basis, as implemented in the Gaussian 03 software (Gaussian, Inc.) [17]. The 324 single point calculations resulted for each sugar in two magnetic shielding maps, that were converted into chemical shift maps $\sigma(C_1,\phi,\psi)$ and $\sigma(C'_1,\phi,\psi)$. The conversion from rough magnetic shielding into a scale comparable to the experimental chemical shift was carried out by simple linear transformation $\sigma = A \cdot S_{calc.} + B$.

*Simulation of the CPMAS spectrum.*

The scientific background leading to the formulation of the simulated $S(\delta)$ CPMAS spectrum is discussed in detail in the next section, and leads to equation (1). Thus, the numerical evaluation of $S(\delta)$ requires an integration of the chemical shift contours in the map $\sigma(C_x,\phi,\psi)$ of a given $C_x$ carbon, properly weighed for the conformational probability density $\chi(\phi,\psi)$.

Since the actual resolution of this map is limited to 20° angular steps in the present work, an interpolation method is needed in order to smooth the integration contour. Several methods have been proposed in the literature [7, 18], including series expansions on a trigonometric basis. In the present work, reliable data have been provided by using simple cubic spline interpolation of the contours, with the advantage of being easily implemented following standard procedures of elementary linear algebra. The simulated CPMAS spectrum $S(\delta)$ was evaluated by sampling a sufficient number of data points (1024 different $\delta$ values), each one corresponding to one numerical integration over an interpolated $\sigma(C_x,\phi,\psi)=\delta$ contour, evaluated by standard Romberg integration method. The gradient term in equation (1) was calculated by finite difference approximation using $\Delta\phi=\Delta\psi=0.01°$ angular steps.

The last step of data processing is the convolution of $S(\delta)$ with a lorentzian function, in order to account for the homogeneous CPMAS linewidth. This was achieved by inverse Fourier Transform of $S(\delta)$ to a free induction decay $S(t)$, then multiplication of $S(t)$ by an exponential function and back Fourier Transform to $S(\delta)$. A typical line broadening imposed in this work was 1 ppm HWHM (half width at half of the maximum intensity).

**Selection of an accurate chemical shift evaluation method**

A first step towards correct modelling of conformational dependence of $^{13}$C isotropic chemical shifts in solid state is to isolate a reliable calculation. Although *ab initio* GIAO strategies have often proven their efficiency and are becoming more and more popular, their spreading in routine applications is still hindered by a considerable computational expense, and other semi-empirical methods have been alternatively proposed [19, 20]. We tested the ability of one semi-empirical and one *ab initio* method to reproduce the isotropic chemical shifts observed for some disaccharides crystalline structures.

Figure 1 shows the correlation between calculated and experimental $^{13}$C NMR shifts obtained either by TNDO/2 (as implemented in the HyperNMR module of the HyperChem Pro 7 software package) or by the DFT/GIAO method at different levels of theory. The calculations of the $^{13}$C isotropic chemical shifts were carried out on the crystallographic molecular structures [16, 21-23]. On Figure 1 (a), it is clear that the overall correlation (dashed line) is poor in case of TNDO/2. The results in (a) can be divided in three main groups : the highest-field resonances (exp. 60-75 ppm), intermediate ones (exp. 70-90 ppm), and the lowest-field resonances (exp. 90-110 ppm). For the two first groups, it is still possible to identify some partial correlation between calculated and experimental shifts (solid lines). This suggests that TNDO/2 can describe to some extent some structural features of the carbons of the pyranose or furanose rings, and of the primary alcohol groups. However, the mean square deviation for the calculated lowest field resonances (encircled), assigned to the carbons linked to the bridging oxygen, is more than 10 ppm. This very high value rules out the TNDO/2 method for investigating this region of the CPMAS spectrum. These observations suggest that empirically adjusted methods are able to account for first order contributions to the chemical shift (nature of the chemical group), but not for second order ones (distortion of the valence orbitals due to bonding geometry), and therefore cannot be employed in conformational studies of disaccharides in the solid state. Figure 1 (b) shows the results obtained with the DFT/GIAO method on the same molecules. The linear regression factors exceed R=0.99 (m.s.d. less than 3 ppm) for all hamiltonians or base length tested, and the slope of the best linear fit ranges from -0.85 to -0.92. The largest deviations are observed for carbons which are implied in intermolecular hydrogen bonds (according to the x-ray structures), which cannot be accounted for in single molecule calculations. It is interesting to note that a moderate base length (3-21+g**) was found sufficient to reproduce correctly the CPMAS spectra of crystalline sugar structures. Finally, the B3PW91/3-21+g** method was chosen, in agreement with previous works on glucose derivatives [24].

# Relating observed chemical shift distributions to structural information on amorphous disaccharides

## Variation of chemical shift

Amorphous forms of disaccharides often display large distributions of isotropic chemical shift in their CPMAS $^{13}$C NMR spectrum, especially for the carbons of the primary alcohol groups and for two carbons involved in the glycosidic bond. The general way of relating the experimental chemical shift distribution assigned to one particular $^{13}$C$_x$ atom to a distribution of molecular conformations in the amorphous (glassy) state first implies the identification of all structural parameters $\Omega=(\theta_1,\theta_2, \ldots,\theta_N)$ (bond angles, bond lengths, dihedral angles…) that significantly influence the magnetic shielding around this $^{13}$C$_x$ nucleus. Spanning this N dimensional parameter space over all the corresponding molecular conformations affords the N dimensional $\sigma(C_x,\theta_1,\theta_2, \ldots,\theta_N)$ chemical shift maps to be constructed for each C$_x$ carbon. For the two carbons implied in a glycosidic bond of disaccharides, it is commonly accepted that only two ($\Omega=(\phi,\psi)$) or three ($\Omega=(\phi,\psi,\omega)$) dihedral angles defining the local geometry of the glycosidic linkage are sufficient to describe the observed chemical shift values (N=2 for all disaccharides except those linked 1-6, for which the third rotation ω gives N=3). Whether this approximation may be satisfactory or not is question of debate, but presently it represents a convenient computational route for complex saccharidic moieties and will be checked out in this section.

For a simple 1-4 linked disaccharide, two 2D chemical shift surfaces $\sigma(C_1,\phi,\psi)$ and $\sigma(C'_4,\phi,\psi)$ should be enough to account for the two observed chemical shift distributions of the non-reducing residue carbon atom $C_1$ and of the reducing residue carbon atom $C'_4$, respectively. Under these circumstances, comparison of CPMAS data may give a significant opportunity to reach a structural information on the local distribution of conformations around the glycosidic bond in the non-crystalline state of sugars. The presence of symmetry in the structure of trehalose (α-ᴅ-glucopyranosyl-(1,1)-α-ᴅ-glucopyranoside, Figure 2), introduces some further simplification, at least as it concerns the amount of calculation. For this disaccharide, only one 2D chemical shift surfaces $\sigma(C_1,\phi,\psi)$, because $C_1$ and $C'_1$ become indistinguishable.

Figure 3 shows the evolution of the magnetic shielding of the $C_1$ as a function of one dihedral angle (ϕ in the figures 4 & 5) and leaving constant the other angle (ψ is set to 60.8°). For all points in figures 3a and 3b, the NMR parameters were calculated using the DFT B3PW91 hamiltonian, and the 3-21+g** basis set. In Figure 3a, the different points refer to calculations carried out on molecular conformations with same fixed ϕ and ψ, the ring geometries being optimized using very different methods, from semi-empirical (AM1) to *ab initio*, and different basis sets (up to 6-311+g**). Hence, different results find their origin either in molecular geometrical parameters other than the conformation of the glycosidic bond, or in the level of theory employed. Only by translating the different curves by an offset adapted to each case, we see that they all collapse into a mastercurve, within a confidence interval of about ±1 ppm. Figure 3b shows a similar evolution of the chemical shift of the $C_1$ as a function of $\phi$. The two curves refer to calculations carried out by using the BIO85 force-field, one constraining the pyranose ring in the rigid geometry of the LMC, the other relaxing all ring parameters. The results clearly show that the dihedral angle $\phi$ is a determinant parameter for the $C_1$ chemical shift (of course, a similar evolution is verified also for the $\psi$ angle). It is also noteworthy that all other ring parameters, which are completely relaxed during the optimizations (whatever the method), have a smaller contribution (less than 1 ppm). This observation justifies the mentioned assumption that the knowledge of $\Omega=(\phi,\psi)$ can be sufficient to account for the $^{13}C$ shifts of $C_1$ (and $C'_1$) of trehalose to a reasonable level of accuracy. Even more important, at least for the glycosidic linkage of trehalose, the results of Figure 3 suggest that the values of $\sigma(C_1,\phi,\psi)$ (and $\sigma(C'_1,\phi,\psi)$) do not seem significantly affected by the force field used to evaluate the molecular geometry. Hence, to a good approximation it should be also possible to calculate independently the $\sigma(\phi,\psi)$ and the $E(\phi,\psi)$ maps. Relying on this observation, the aim of this section is to demonstrate that the $\sigma(\phi,\psi)$ map, calculated with a given but sufficiently accurate force field, can be considered as universal and can be therefore used to compare the ability of different conformational energy landscape models $E(\phi,\psi)$ to fit the experimental CPMAS data.

**Numerical procedure for calculating chemical shift distributions**

The procedure for calculating the CPMAS spectrum associated with a given carbon atom $C_x$ is detailed in this section. First, a Ramachandran map $E(\phi,\psi)$ is calculated over 324 relaxed disaccharide conformations in a given molecular mechanics force field. Second, the corresponding $\sigma(C_x,\phi,\psi)$ map is evaluated over the same 324 conformations. The choice of the DFT B3PW91/3-21+g** Hamiltonian/basis provided the computation of this map in about 10 CPU days for trehalose (and less than three CPU weeks for other disaccharides). The two maps of $E(\phi,\psi)$ and $\sigma(C_x,\phi,\psi)$ are shown in Figure 4 for trehalose. As discussed in the previous section, the chemical shift surface will be in the following considered as independent on the force-field and method chosen for relaxing the conformations.

The energy surface (Figure 4a) compares well with other similar maps for trehalose [5, 25]. For sake of comparison, however, it should be necessary to take into account that literature works often refer to solvated trehalose and specifically address attention to the effect of solvent (water) on the conformational energy map [26]. Nonetheless, it can be easily recognized that the most recent potential of mean force surface [27] and the previous quantum mechanical calculations at high level of theory [5, 25] show that the conformation of the global minimum structure for the isolated trehalose molecule resides in the region of $50° < \phi,\psi <100°$. Both these works and other previous results also indicate the existence of other relative minima that may become relevant under specific intermolecular interactions (either with solvent or with other trehalose molecules).

The DFT chemical shift surface is very similar to the one obtained with a Hartree-Fock method [5], although the latter calculations refer to a model molecule (tetrahydropyran-dimer) resembling trehalose. For this analog, also the energy surface calculated in ref [5] presents substantial similarities with that shown in Figure 4. These observations justify the correctness of the choice of Grandinetti et al. in favour of a simpler molecular analogue of trehalose and furthemore confirm that local interactions dominate across the glycosidic linkage. More difficult appears the visual comparison that can be made with the chemical shift results reported by Moyna and coworkers in form of three-dimensional plots [18, 28]. However, it is enough clear from all the above discussed data that the gross features in the chemical shift of the populated region do not show significant changes, a fact that supports the independence of the $\sigma(C_1,\phi,\psi)$ map from other structural details.

In the amorphous state, quenched disaccharide conformations coexist, distributed over the $(\phi,\psi)$ space according to an occupation probability density $\chi(\phi,\psi)$. This function alone entirely carries all the information that, in principle, can be extracted from experimental CPMAS data on the structurally disordered glass. In amorphous forms with a glassy character, the function $\chi(\phi,\psi)$ can be in first approximation considered reminiscent of the conformational space that is dynamically explored in the undercooled liquid state at temperatures above $T_g$. In other words, it is assumed that the ergodic equivalence exists between the time average of conformational liquid states and the spatial average of amorphous state. Obviously, a Boltzmann statistic is the simplest tool to relate $\chi(\phi,\psi)$ to a model of conformational energy landscape $E(\phi,\psi)$. Such an equation is dependent on $E(\phi,\psi)$ (single molecule property), assuming that the internal energy landscape is explored thermally without intermolecular correlation. In order to check this point, at least partially, other simple models for $\chi(\phi,\psi)$ will be also discussed in this paper. It is however important to note that the procedure described in this section can also be applied to more realistic models for both $\chi(\phi,\psi)$ and $E(\phi,\psi)$, that could in the same manner be confronted to experimental results.

The inhomogeneous broadening of the NMR lineshape resulting from quenched conformational disorder in glassy disaccharides can be described by a spectral function of the chemical shift $S(\delta)$. This function reflects the spectral density of the $^{13}C$ resonances lying in $\varepsilon=[\delta, \delta+d\delta]$, and can be evaluated as the angular density of states resonating in $\varepsilon$ weighed by the occupation probability density $\chi(\phi,\psi)$ of these states, and can be written :

$$S(\delta) = \oint_{\sigma(\phi,\psi)=\delta} \frac{\chi(\phi,\psi)}{\left|\vec{\nabla}\sigma(\phi,\psi)\right|_{\sigma=\delta}} d\phi d\psi \tag{1}$$

The integral in equation (1) runs along the contour of constant chemical shift $\sigma = \delta$ in the map $\sigma(C_x,\phi,\psi)$. Discrete approximations of this equation have been used in the literature either for solid state studies [5] or in time average calculations for solution studies [7, 18]. It is however to point out that only the presence of the gradient term of the chemical shift surface can induce discontinuities in $S(\delta)$ for angular contributions where this gradient vanishes. Such spectral discontinuities are often observed in CPMAS spectra in presence of inhomogeneous broadening, and provide basic features easily measurable. For instance, equation (1) can be directly used for calculating chemical shift (CS) anisotropy powder patterns, by replacing the conformational $(\phi,\psi)$ angles by the Euler angles $(\theta,\phi)$ locating the orientation of the CS tensor, and $\chi(\phi,\psi)=\sin(\theta)$ (the Jacobian of the transformation to spherical coordinates).

**Energy landscape models below $T_g$ in amorphous trehalose**

Provided correct chemical shift maps are known, the comparison between simulated CPMAS spectra obtained by equation (1) and observed NMR data provides a test whether a given model $\chi(\phi,\psi)$ of occupied conformations in the glass is compatible with the experiment. The maximum value of such a compatible $\chi(\phi,\psi)$ population is a major information for it represents the most probable molecular conformation (LMC) in the solid. The position $(\phi_0,\psi_0)$ of this maximum should therefore coincide with an energy minimum of the energy landscape of the glass. Also the detailed features of $\chi(\phi,\psi)$ are important, inasmuch can be related to the shape of the energy basin around $(\phi_0,\psi_0)$ through a partition function depending on $E(\phi,\psi)$ (Boltzmann statistic for instance). The global method exposed in the preceding sections can in fact be used to confront energy landscape models to the experiments. First, the model can be tested in a structural approach below $T_g$, where chemical shift distributions correspond to static conformational distributions. Second, the energy landscape concept is intrinsically a dynamical model, that can also be tested at temperatures above $T_g$, where the experimental lineshapes begin to be dynamically averaged. Note that these conditions ($T \geq T_g$) can be matched either by increasing temperature or by increasing molecular mobility through the addition of a proper plasticizer. Under such circumstances, fast reorientation is expected to completely average the NMR lineshapes. This fast-reorientation limit has already been subject to abundant literature coupling NMR and simulations (see e.g. [29, 30] and references therein).

The energetic model describing the partition of conformations in a glassy solid is restricted to intramolecular degrees of freedom. For simple disaccharides, it has already been discussed in the previous sections that this approach can be reduced to a two dimensional phase space, leading to the definition of a Ramachandran map $E(\phi,\psi)$. Within this approach, amorphous disaccharides are described as the sum of individual molecules in different conformations, and intermolecular correlations are neglected. It has been shown [5] that this approach can lead to a good description of the observed NMR spectrum in case of an $\alpha$-trehalose analog, when the $E(\phi,\psi)$ surface is calculated with the same *ab initio* method as for the $\sigma(C_x,\phi,\psi)$ chemical shift surface. The use of a simple molecule as a model for trehalose may have been dictated by the necessity of limiting the computational time, but also has been based upon the likely hypothesis of the negligible effects of long-range molecular details. Still, this all *ab initio* method is so much time consuming that it would be ruled out for routine studies on more complex molecules. In this section, additional results for trehalose as a test molecule are presented, by calculating the $E(\phi,\psi)$ map with more simple molecular mechanics force fields, and keeping an *ab initio* standard for NMR calculations. Choosing molecular mechanics allows the E($\phi,\psi$) map to be evaluated within a couple of hours only on standard machines.

Figure 5 compares the simulated $^{13}$C CPMAS spectra of trehalose calculated according to equation (1) with different choices for the $\chi(\phi,\psi)$ conformational population. Figure 5 (a) presents simulations using Boltzmann partitions on E($\phi,\psi$) maps calculated by some popular force fields. It is clear that different parameters sets (i.e. different E($\phi,\psi$) maps) produce different results (as repeatedly reported in literature), and that only the parameters BIO85 seem to provide a suitable NMR lineshape simulation. However, scrutiny of the Ramachandran maps obtained with these force fields (see Figure 4a) reveals that the more relevant difference occurs in the position ($\phi_0,\psi_0$) of the lowest energy minimum (most probable molecular conformation - LMC). The position of the LMC for the BIO85 parameters is near (60°,60°), which is very close to the crystalline conformations of trehalose. Comparatively, the LMC positions found with CHARMM27 and AMBER99 force-field parameters are close to (90°,90°) and (60°,60°), respectively. In the CHARMM27 force-field, the preferred conformational population gets closer to gauche conformations, which is evidently not consistent with the observed CPMAS spectrum. This conformational aspect of trehalose has already been long debated, and our observations are in agreement with the *ab initio* investigation of French *et al.* [25]. One should, however, also reminds the suitability of the above reported force fields to simulate carbohydrates in a self-made matrix without solvent. In order to confirm that the origin of the disagreement between simulated and experimental CPMAS spectra in the case CHARMM27 force field is mainly due to the shift of the LMC position, simplistic gaussian $\chi(\phi,\psi)$ functions of same isotropic width have been used. Two gaussian partitions with the same amplitude (HWHM) are shown in Figure 5 (b) : the first function is centered close to the lowest energy minimum obtained with the BIO85 parameters, whereas the second function is diagonally shifted from that position (a shift of 20° has been applied to both torsional angles). It turns out clearly that the gaussian distribution centered on (70°,70°) gives the best fit of the experimental lineshape. The small shift from this optimal position is sufficient to induce extra features in the simulated spectrum, that are unambiguously inconsistent by simple comparison with the experimental lineshape. It can be concluded that either a Boltzmann partition on the BIO85 map or a simple gaussian give equally acceptable results. This result has many implicit meanings: on one side, it give confidence to the calculation that match an overall correct spatial distribution of most probable conformations; on the other side, it seems not possible to scrutinize clearly the detail of the expected realistic shape of the conformational distribution for this molecular system.

Another information that is reached through this simulation is the average extent of the $\chi(\phi,\psi)$ distribution in the $(\phi,\psi)$ conformational space. Either gaussian or Boltzmann (BIO85) approaches agree that the HWHM of the conformational distribution is about 50°. This value is expected to be independent on $E(\phi,\psi)$, but highly dependent on $\sigma(C_1/C'_1,\phi,\psi)$. Therefore, a correct evaluation of the actual conformational range explored in the glass is submitted to a very accurate definition of the chemical shift map. An additional comment seems rather necessary here. The single symmetric energy basin occurring in trehalose is rather uncommon in biomolecules; other sugar dimers and especially peptides are common to explore several distinct minima, with the result of increasing the distribution of conformations exhibiting likely different chemical shifts. Indeed, and very important, for molecules showing several minima displaced over the whole conformational angles, a simple rotational isomeric state (RIS) approximation could probably suffice.

Although clear evidence is given in section 10 that the DFT B3PW91/3-21+g** basis is sufficient to reproduce relative carbon resonances in different chemical environments, no experimental criterion does exist up to now that would validate that this basis is able to correctly describe the second order conformational dependence of the chemical shift. The range of 50° found in our work is significantly greater than the one proposed by Zhang *et al.* [5]. This probably overestimated value suggests that the B3PW91/3-21+g** basis underestimates the ($\phi,\psi$) dependence of the chemical shift. Indeed, more systematic studies of the influence of the *ab initio* basis length should be undertaken, at the expense of CPU time. The temperature value that must be injected in the Boltzmann law linking the BIO85 E($\phi,\psi$) and $\chi(\phi,\psi)$ is also very high (about 1800 K), and carries poor physical meaning. This suggests that the potential basin is not correctly described by molecular mechanics (overestimation of potential barriers), or that intermolecular correlations, that are not considered in the presented simple approach, result in the stabilisation of relatively high energy conformations in the glass [5]. Apart from these reserves, the conclusion can be safely reached that the coupling of molecular mechanics to GIAO calculations provides a precise, unambiguous and model independent determination of the most probable conformation of trehalose in a glassy solid state form, and that it can be used to selectively discriminate between different energy landscape models.

## Conclusion

The aim of this article was to work out and exploit the correlation between simulated chemical shift surfaces and the distribution of the conformational disorder in glassy solids, such as disaccharides, in relation with conformational energy landscape models. The working hypothesis was based on the possibility that the chemical shift surface $\sigma(C_x,\phi,\psi)$ for disaccharides is primarily dependent on the dihedral angles $\phi$ and $\psi$ defining the geometry of the glycosidic bond.

The present results show that:

1) There is a universality character conferred to the chemical shift surface, making it almost independent on the method chosen for optimising the molecular conformations. It also implies that *ab initio* methods are not always necessary for defining the conformational potential basin. Indeed, the calculations based on the coupling between molecular mechanics and DFT/GIAO provide reliable results within a reasonable amount of CPU time.

2) Furthermore, it has been shown that a rigorous numerical treatment must be followed in order to properly account for the observed discontinuities of the CPMAS NMR spectrum. This requires a correct integration of the extrema and saddle points of the chemical shift surface, through a gradient term in the simulated spectral density $S(\delta)$. A detailed comparison of the results obtained by different elementary structural models is presented, showing that the comparison between simulated and experimental spectra of trehalose is highly discriminative of the most probable molecular conformation in the glassy state. This information is of considerable importance for a better understanding of the quenching of the intramolecular degrees of freedom below the glass transition temperature. On the other hand, the detailed shape of the energy profile around the most probable conformation seems to have less influence on the simulated NMR lineshape.

3) As a further conclusion, the use of $^{13}$C CPMAS spectra simulation based on chemical shift surfaces provides a unique opportunity to confront energy landscape models of the glass to experimental data. In particular, dynamical models should be compared to the partially motionally averaged NMR spectra near the glass transition temperature. Application of the method detailed in the present paper to other disaccharides with different bond or ring types is currently in progress.

## Acknowledgments

A.C. is grateful to University Lille I for Visiting Professorship and to hosting LDSMM for hospitality.

## References

1. Venable, R.M., F. Delaglio, S.E. Norris, and D.I. Freedberg, *The utility of residual dipolar couplings in detecting motion in carbohydrates: application to sucrose.* Carbohydrate Research, 2005. **340**(5): p. 863-874.

2. Hemmingsen, L., D.E. Madsen, A.L. Esbensen, L. Olsen, and S.B. Engelsen, *Evaluation of carbohydrate molecular mechanical force fields by quantum mechanical calculations.* Carbohydrate Research, 2004. **339**(5): p. 937-948.

3. Perez, S., A. Imberty, S.B. Engelsen, J. Gruza, K. Mazeau, J. Jimenez-Barbero, A. Poveda, J.-F. Espinosa, B.P. van Eyck, and G. Johnson, *A comparison and chemometric analysis of several molecular mechanics force fields and parameter sets applied to carbohydrates.* Carbohydrate Research, 1998. **314**(3-4): p. 141-155.

4. Furlan, S., G. La Penna, A. Perico, and A. Cesaro, *Hyaluronan chain conformation and dynamics.* Carbohydrate Research, 2005. **340**(5): p. 959-970.

5. Zhang, P., A.N. Klymachyov, S. Brown, J.G. Ellington, and P.J. Grandinetti, *Solid-state NMR investigations of the glycosidic linkage in α-α' trehalose.* Solid State Nuclear Magnetic Resonance, 1998. **12**(4): p. 221-225.

6. Swalina, C.W., R.J. Zauhar, M.J. DeGrazia, and G. Moyna, *Derivation of 13C chemical shift surfaces for the anomeric carbons of oligosaccharides and* *glycopeptides using ab initio methodology.* Journal of Biomolecular NMR, 2001. **21**: p. 49-61.

7. O'Brien, E.P. and G. Moyna, *Use of 13C chemical shift surfaces in the study of carbohydrate conformation. Application to cyclomaltooligosaccharides (cyclodextrins) in the solid state and in solution.* Carbohydrate Research, 2004. **339**(1): p. 87-96.

8. Lefort, R., A. De Gusseme, J.-F. Willart, F. Danède, and M. Descamps, *Solid state NMR and DSC methods for quantifying the amorphous content in solid dosage forms: an application to ball-milling of trehalose.* International Journal of Pharmaceutics, 2004. **280**(1-2): p. 209-219.

9. Paris, M., H. Bizot, J. Emery, J.Y. Buzare, and A. Buleon, *NMR local range investigations in amorphous starchy substrates I. Structural heterogeneity probed by 13C CP-MAS NMR.* International Journal of Biological Macromolecules, 2001. **29**(2): p. 127-136.

10. Paris, M., H. Bizot, J. Emery, J.Y. Buzare, and A. Buleon, *NMR local range investigations in amorphous starchy substrates: II-Dynamical heterogeneity probed by 1H/13C magnetization transfer and 2D WISE solid state NMR.* International Journal of Biological Macromolecules, 2001. **29**(2): p. 137-143.

11. Maunu, S.L., *NMR studies of wood and wood products.* Progress in Nuclear Magnetic Resonance Spectroscopy, 2002. **40**(2): p. 151-174.

12. Tishmack, P.A., D.E. Bugay, and S.R. Byrn, *Solid-state nuclear magnetic resonance spectroscopy - pharmaceutical applications.* Journal of Pharmaceutical Sciences, 2003. **92**(3): p. 441-474.

13. Reiher, W.E., III., *Ph. D. Thesis ; Theoretical studies of hydrogen bonding*, in *Chemistry department*. 1985, Harvard University: Cambridge, MA.

14. Foloppe, N. and J. MacKerell, A., D., *All-Atom Empirical Force Field for Nucleic Acids : 1) Parameter Optimization Based on Small Molecule and Condensed Phase Macromolecular Target Data.* Journal of Computational Chemistry, 2000. **21**: p. 26-104.

15. Cornell, W.D., P. Cieplak, C.I. Bayly, I.R. Gould, K.M.J. Merz, D.M. Ferguson, D.C. Spellmeyer, T. Fox, J.W. Caldwell, and P.A. Kollman, *A Second Generation Force Field for the Simulation of Proteins, Nucleic Acids, and Organic Molecules.* Journal of the American Chemical Society, 1995. **117**: p. 5179-5197.

16. Jeffrey, G.A. and R. Nanni, *The Crystal Structure of Anhydrous α,α-Trehalose at -150°.* Carbohydrate Research, 1985. **137**: p. 21-30.

17. M. J. Frisch, G. W. Trucks, H. B. Schlegel, G. E. Scuseria, M. A. Robb, J. R. Cheeseman, J. A. Montgomery, T.V. Jr., K. N. Kudin, J. C. Burant, J. M. Millam, S. S. Iyengar, J. Tomasi, V. Barone, B. Mennucci, M. Cossi, G. Scalmani, N. Rega, G. A. Petersson, H. Nakatsuji, M. Hada, M. Ehara, K. Toyota, R. Fukuda, J. Hasegawa, M. Ishida, T. Nakajima, Y. Honda, O. Kitao, H. Nakai, M. Klene, X. Li, J. E. Knox, H. P. Hratchian, J. B. Cross, C. Adamo, J. Jaramillo, R. Gomperts, R. E. Stratmann, O. Yazyev, A. J. Austin, R. Cammi, C. Pomelli, J. W. Ochterski, P. Y. Ayala, K. Morokuma, G. A. Voth, P. Salvador, J. J. Dannenberg, V. G. Zakrzewski, S. Dapprich, A. D. Daniels, M. C. Strain, O. Farkas, D. K. Malick, A. D. Rabuck, K. Raghavachari, J. B. Foresman, J. V. Ortiz, Q. Cui, A. G. Baboul, S. Clifford, J. Cioslowski, B. B. Stefanov, G. Liu, A. Liashenko, P. Piskorz, I. Komaromi, R. L. Martin, D. J. Fox, T. Keith, M. A. Al-Laham, C. Y. Peng, A. Nanayakkara, M. Challacombe, P. M. W. Gill, B. Johnson, W. Chen, M. W. Wong, C. Gonzalez, and J.A. Pople, *Gaussian 03, Revision B.05,*, I. Gaussian, Editor. 2003: Pittsburgh PA.

18. Swalina, C.W., R.J. Zauhar, M.J. DeGrazia, and G. Moyna, *Derivation of 13C chemical shift surfaces for the anomeric carbons of oligosaccharides and glycopeptides using ab initio methodology.* Journal of Biomolecular NMR, 2001. **21**(1): p. 49-61.

19. Young, D.C., *Semiempirical Methods*, in *Computational Chemistry*. 2002, Wiley: New York.

20. Heine, T. and G. Seifert, *Semiempirical Methods for the Calculation of NMR Chemical Shifts*, in *Calculation of NMR and EPR Parameters*, Prof. Dr. Martin Kaupp, Dr. Michael Bühl, and D.V.G. Malkin, Editors. 2004, Wiley.

21. Brown, G.M., D.C. Rohrer, and B. Berking, *The Crystal Structure of α,α-Trehalose Dihydrate from Three Independent X-ray Determinations.* Acta Crystallographica, 1972. **B28**: p. 3145-3158.

22. Brown, G.M. and H.A. Levy, *Further Refinement of the Structure of Sucrose Based on Neutron-Diffraction Data.* Acta Crystallographica, 1973. **B29**: p. 790-797.

23. Fries, D.C., S.T. Rao, and M. Sundralingam, *Structural Chemistry of Carbohydrates. III. Crystal and Molecular Structure of 4-O-β-D-Galactopyranosyl-α-D-glucopyranose Monohydrate (α-Lactose Monohydrate).* Acta Crystallographica, 1971. **B27**: p. 994.

24. Kupka, T., G. Pasterna, P. Lodowski, and W. Szeja, *GIAO-DFT prediction of accurate NMR parameters in selected glucose derivatives.* Magnetic Resonance in Chemistry, 1999. **37**(6): p. 421-426.

25. French, A.D., G.P. Johnson, A.-M. Kelterer, M.K. Dowd, and C.J. Cramer, *Quantum Mechanics Studies of the Intrinsic Conformation of Trehalose.* Journal of Physical Chemistry A, 2002. **106**(19): p. 4988-4997.

26. Liu, Q., R.K. Schmidt, B. Teo, P.A. Karplus, and J.W. Brady, *Molecular Dynamics Studies of the Hydration of α,α-Trehalose.* J. Am. Chem. Soc., 1997. **119**(33): p. 7851-7862.

27. Kuttel, M.M. and K.J. Naidoo, *Ramachandran free-energy surfaces for disaccharides: trehalose, a case study.* Carbohydrate Research, 2005. **340**(5): p. 875-879.

28. Sergeyev, I. and G. Moyna, *Determination of the three-dimensional structure of oligosaccharides in the solid state from experimental 13C NMR data and ab initio chemical shift surfaces.* Carbohydrate Research, 2005. **340**(6): p. 1165-1174.

29. *Calculation of NMR and EPR Parameters*, ed. P.D.M. Kaupp, D.M. Bühl, and D.V.G. Malkin. 2004, New-York: Wiley-VCH Verlag GmbH & Co. KGaA. 603.

30. Forsyth, D.A. and A.B. Sebag, *Computed 13C NMR Chemical Shifts via Empirically Scaled GIAO Shieldings and Molecular Mechanics Geometries. Conformation and Configuration from 13C Shifts.* Journal of the American Chemical Society, 1997. **119**(40): p. 9483-9494.

**Figure Captions**

Figure 1 : Calculated ($\sigma_{calc}$) isotropic 13C chemical shift (a) or magnetic shielding (b) of crystalline trehalose dihydrate, anhydrous trehalose, lactose monohydrate and sucrose versus experimental chemical shifts ($\sigma_{exp}$). The results for all carbons of the four dimeric sugars are presented together. Calculations were carried out on isolated sugar molecule, with atomic coordinates taken from crystallographic data and hydrogen atoms placed at positions corresponding to energy minima of the BIO85 force field: (a) Semi-empirical TNDO method; (b) ab initio density functional theory with GIAO method at levels b3lyp/6-31g+(d,2p) (●), b3lyp/3-21+g** (○), and b3pw91/3-21+g** (▼).

Figure 2 : Snapshot of a trehalose conformation showing the relevant angular parameters and the symmetry occurring in this dimeric sugar (in this conformation, angles are set to $\phi = \psi = 60°$. Hydrogen atoms have been omitted for clarity.

Figure 3 : (a) GIAO Magnetic shielding calculated at the B3PW91//3-21g+** level of theory, on different molecular structures of trehalose. Except for (■), these structures were obtained by fixing the glycosidic dihedral angles to ($\phi = x$ ; $\psi = 60.8°$), then optimizing the other degrees of freedom using : AM1 (●), B3LYP//3-21+g** (○), B3PW91//6-311g+** (▼), B3PW91/3-21+g** (△). For (■), the structure was first fully optimized using B3PW91//3-21+g**, then kept rigid after fixing ($\phi = x$ ; $\psi = 60.8°$) prior to the NMR calculation. (b) GIAO Magnetic shielding calculated at the B3PW91//3-

21g+** level of theory, on different molecular structures of trehalose. First, the structure was fully optimized using the BIO85 empirical force field, then either kept rigid after fixing ($\phi = x$ ; $\psi = 60.8°$) (○), or reoptimized using the same force field for each couple of values ($\phi = x$ ; $\psi = 60.8°$) (●).

Figure 4 : (a) : Adiabatic Ramachandran map $E(\phi,\psi)$ of trehalose calculated with the CHARMM-type BIO85 force field parameters. The fluctuations of the calculations have been averaged following the symmetry of the trehalose molecule: $E(\phi,\psi) = \frac{1}{2}\left[E_{calc.}(\phi,\psi) + E_{calc.}^{T}(\phi,\psi)\right]$. The lowest energy was found for $\phi=\psi=80°$ and was conventionally chosen for origin E=0. (b): $^{13}C$ isotropic chemical shift of trehalose $C_1$ carbon calculated on the same conformations as (a) by GIAO method on a DFT b3pw91/3-21+g** basis. A similar average as in (a) was applied, using the symmetry property $\sigma(C_1,\phi,\psi) = \sigma(C'_1,\phi,\psi)^T$.

Figure 5 : Best fit simulations of the glycosidic region of the $^{13}C$ CPMAS spectrum of trehalose compared to the experiment (thin solid line). The simulations were carried out according to equation (1). Upper line (a) : $\chi(\phi,\psi) = Exp\left(\frac{E(\phi,\psi)}{R.T}\right)$, with $E(\phi,\psi)$ calculated with the following different force fields : BIO85 (bold solid line) and T=1800 K, CHARMM27 (dashed line) and T=600 K or AMBER99 (dotted line) and T=750 K. Lower line (b) : $\chi(\phi,\psi) = Exp\left(-\frac{(\phi-\phi_0)^2 + (\psi-\psi_0)^2}{2\sigma^2}\right)$, with $\sigma=50°$, and $(\phi_0,\psi_0)=(70°,70°)$ or (90°,90°), respectively.

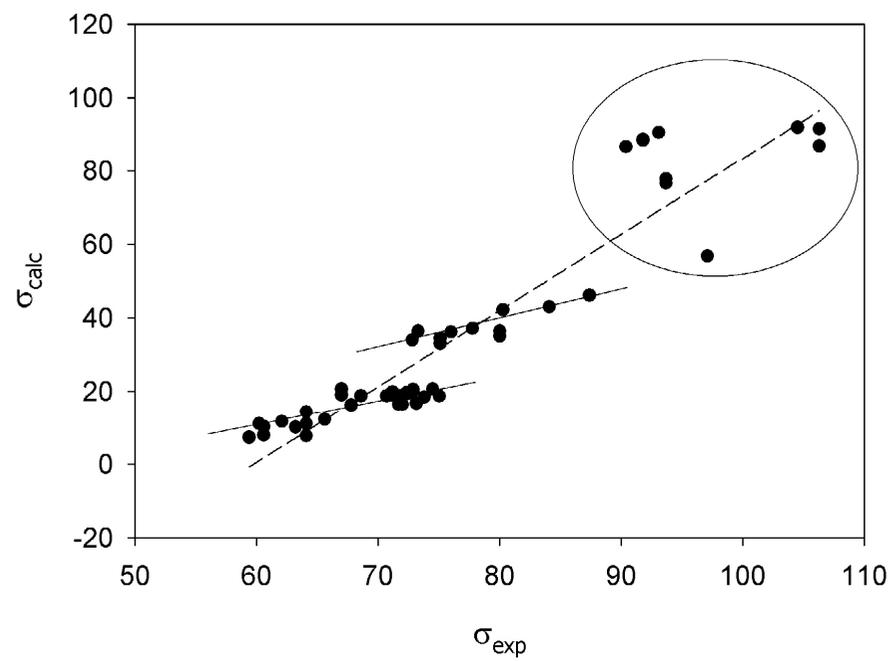 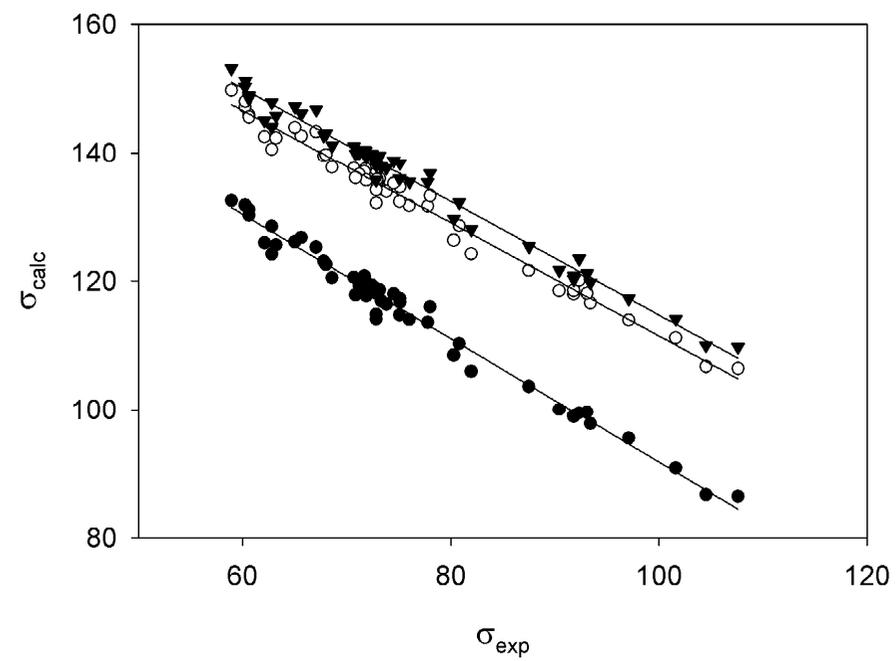

**Figure 1**

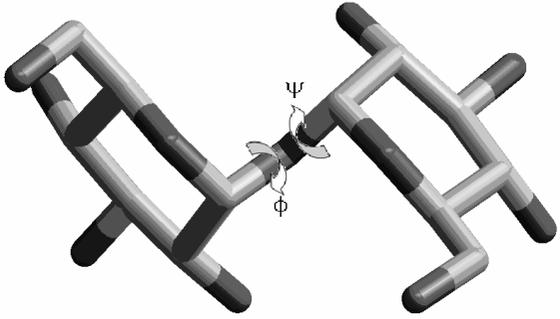

**Figure 2**

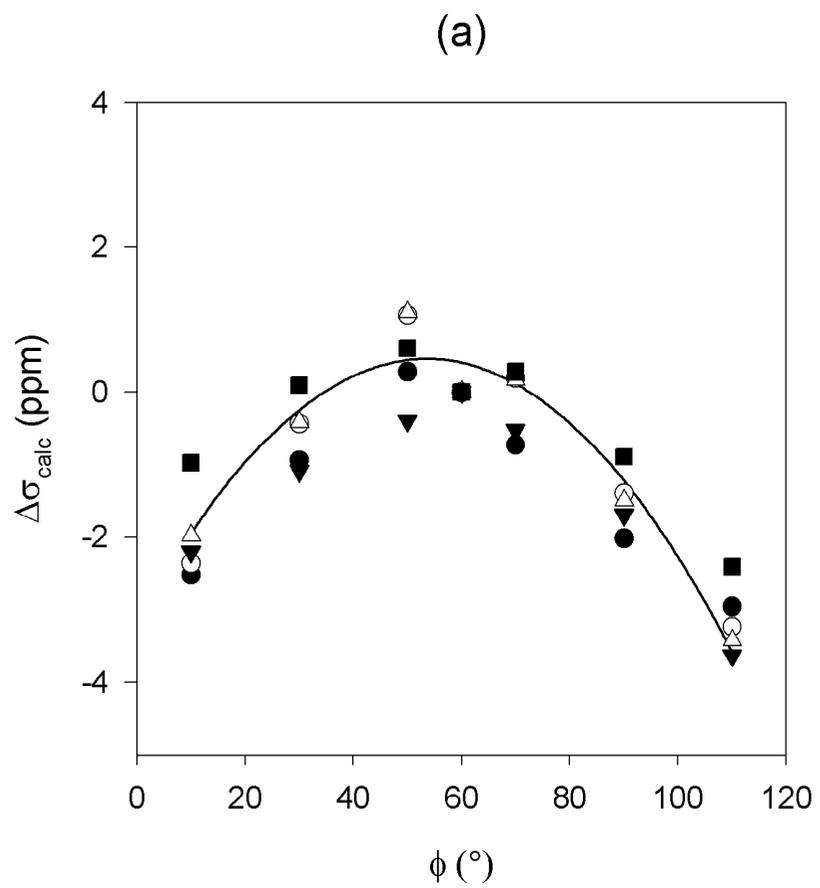 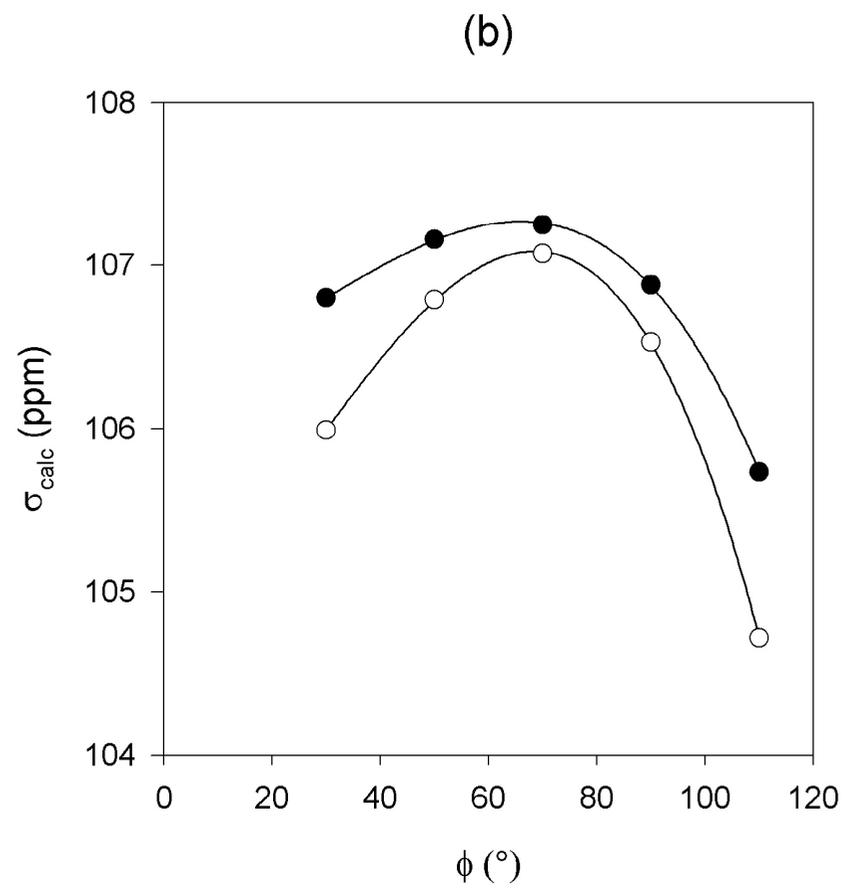

**Figure 3**

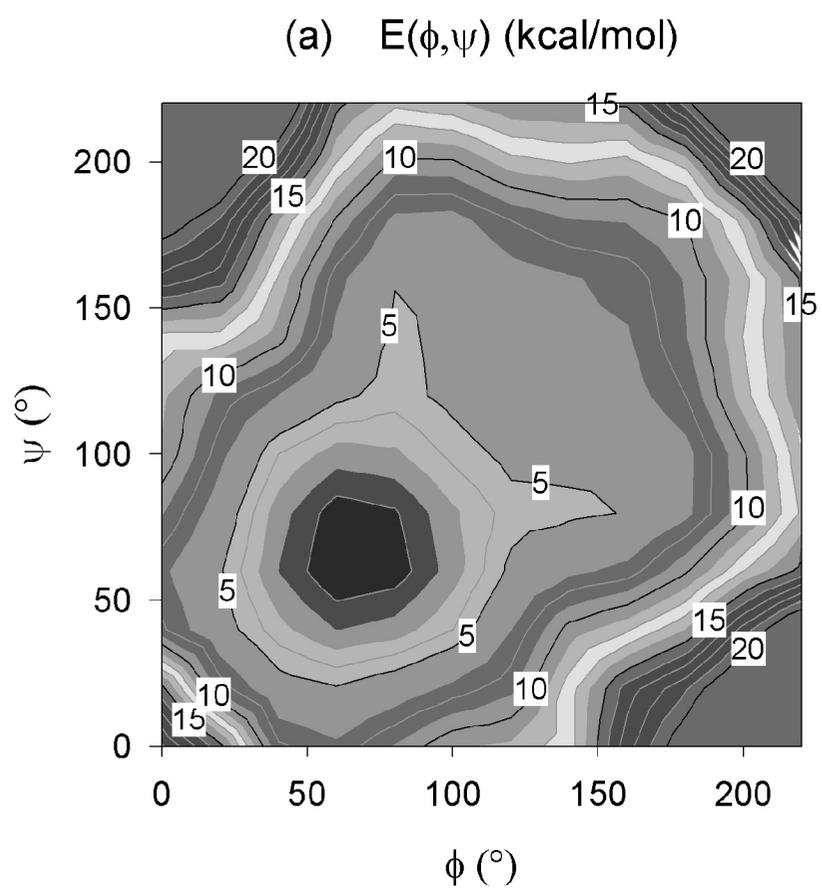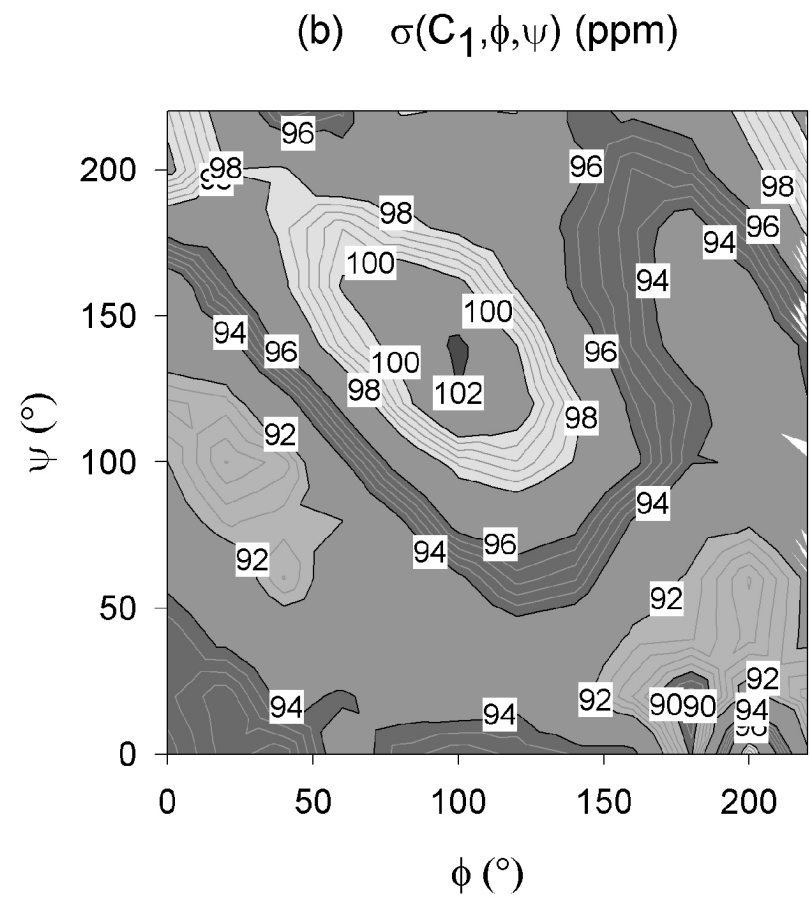

**Figure 4**

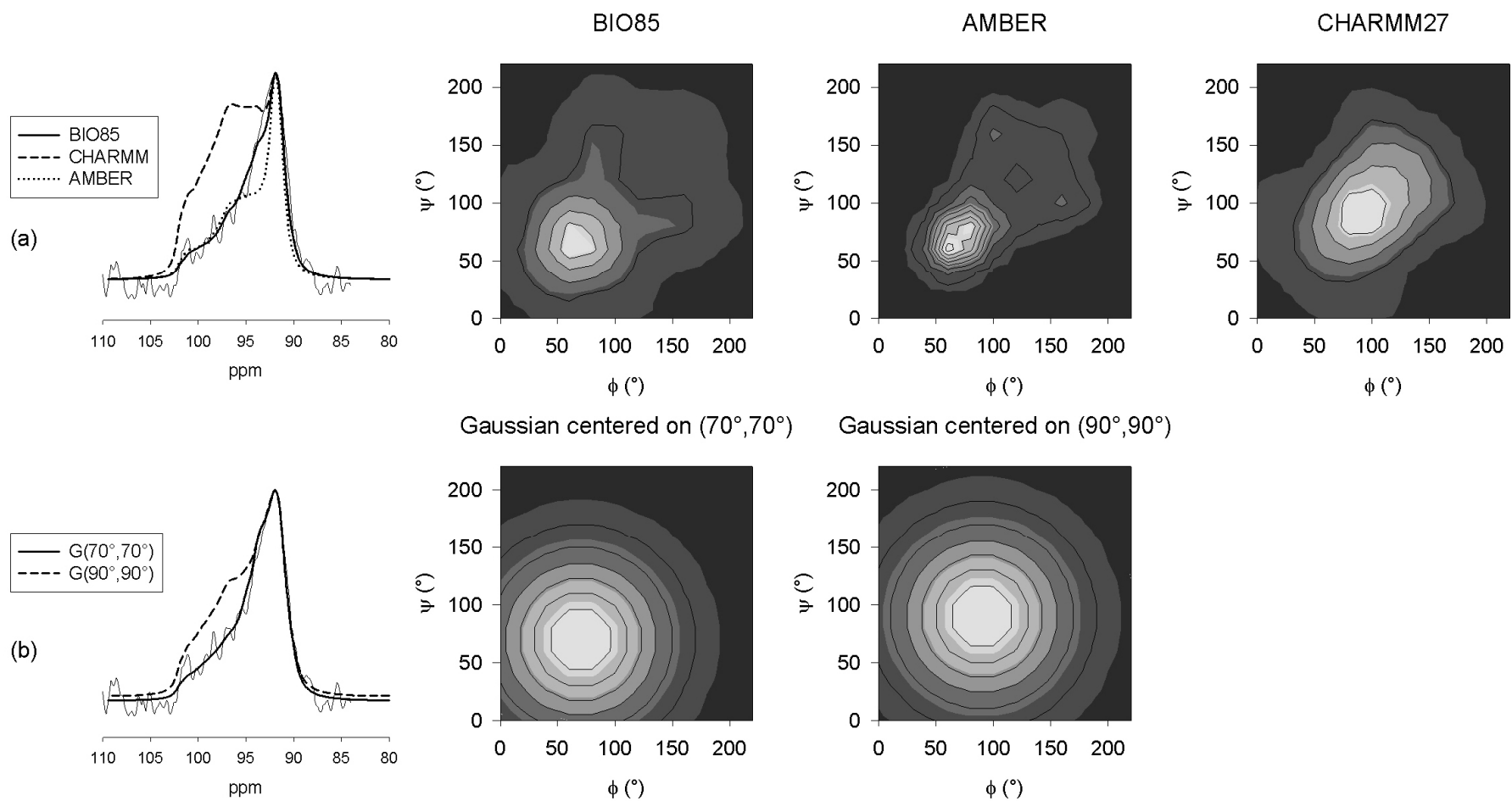

**Figure 5**